\documentclass[]{raa}
\usepackage{}            % referee version: for submission
\input{psfig.sty}
\input{epsf.sty}
\usepackage{graphicx,times}
\usepackage{natbib}

\def\beq{\begin{equation}}
\def\enq{\end{equation}}
\def\ms{$M_\odot$}

\begin{document}

   \title{Temporal Variations And Spectral Properties of Be/X-ray Pulsar GRO J1008-57 Studied by INTEGRAL}

   \volnopage{Vol.0 (201x) No.0, 000--000}      %%preserved for Editor. DOn't remove!
   \setcounter{page}{1}          %%starting page, preserved for Editor. DOn't remove!

   \author{Wei Wang }
%% Here is an example of three authors come from different institutes.
%% For single author or all the authors from an institute, use "\inst{}" only

   \institute{National Astronomical Observatories, Chinese Academy of Sciences,
             Beijing 100012, China; {\it wangwei@bao.ac.cn}}
   \date{Received~~2013 month day; accepted~~2013~~month day}

\abstract{
The spin period variations and hard X-ray spectral properties of Be/X-ray pulsar GRO J1008-57 are systematically studied with INTEGRAL observations during two outbursts in 2004 June and 2009 March. The pulsation periods at $\sim 93.66$ s in 2004 and at $\sim 93.73$ s in 2009 are determined, implying a long-term spin period variation. Pulse profiles of GRO J1008-57 during outbursts are strongly energy dependent: double-peak profile in soft bands of 3 -- 7 keV and single-peak profile in hard X-rays above 7 keV. Combined with previous measurements, we find that GRO J1008-57 has undergone a spin-down trend from 1993 -- 2009 with a rate of $\sim 4.1\times 10^{-5}$ s day$^{-1}$, and might transfer into a spin-up trend after 2009. The hard X-ray spectrum from 3 -- 100 keV of GRO J1008-57 during outbursts can be described by a power-law plus exponential high energy cutoff with a typical photon index of $\Gamma\sim 1.4$ and cutoff energies of $\sim 23-29$ keV. We find a relatively soft spectrum in early phase of the 2009 outburst with cutoff energy $\sim 13$ keV. INTEGRAL also detected GRO J1008-57 in quiescence just before the 2009 outburst. The quiescent spectrum from 3 -- 100 keV is well fitted by a single power-law of $\Gamma\sim 2.1$. Above a hard X-ray flux of $\sim 10^{-9}$ erg cm$^{-2}$ s$^{-1}$, the spectra of GRO J1008-57 during outbursts need an enhanced hydrogen absorption of column density $\sim 6\times 10^{22}$ cm$^{-2}$. The observed dip-like pulse profile of GRO J1008-57 in soft X-ray bands should be caused by this intrinsic absorption. Around the outburst peaks, a possible cyclotron resonance scattering feature at $\sim 74$ keV is detected in the spectra of GRO J1008-57 which is consistent with the report feature by MAXI/GSC observations, making the source as the neutron star of highest magnetic field ($\sim 6.6\times 10^{12}$ G) in known accreting X-ray pulsars. This marginal feature is supported by that the present detections in GRO J1008-57 follow the correlation between fundamental line energies and cutoff energies in accreting X-ray pulsars. Finally we discovered two modulations periods at $\sim 124.38$ days and $\sim 248.78$ days using RXTE/ASM light curves of GRO J1008-57. Two flare peaks appearing in the folded light curve have different spectral properties. The normal outburst lasting 0.1 orbital phase has a hard spectrum and cannot be significantly detected below 3 keV. The second flare lasting ten days shows a very soft spectrum without significant detections above 5 keV. GRO J1008-57 is a good candidate of accreting systems with an equatorial circumstellar disc around the companion star. The neutron star passing the disc of the Be star near periastron and aphelion produces two X-ray flares. Lack of soft X-ray emissions in outbursts should be caused by the enhanced hydrogen absorption near the periastron passage. Soft spectral origins in the second flares still need further detailed studies with soft X-ray spectroscopies. }

\authorrunning{W. Wang }            %author_head in even pages
   \titlerunning{Spin and spectral properties of GRO J1008-57}

\maketitle

\section{Introduction}

The Be/X-ray pulsar GRO J1008-57 was discovered by the BATSE experiment aboard the Compton Gamma Ray Observatory (CGRO) during the 1993 outburst (Stollberg et al. 1993). This source shows the periodic outbursts detected by different X-ray missions in last twenty years. These X-ray outburst activities belong to the Type I outbursts which should be separated by its orbital period. In the BATSE era, the outbursts were detected during 1993 August, 1994 March, 1994 November, and 1996 March, occurring at multiples of $\sim 248$ days, which indicates that this may be the orbital period of the system (Bildsten et al. 1997). Shrader et al. (1999) analyzed the light curve collected by three years of RXTE/ASM monitoring data, suggesting an orbital period of $\sim 136.5$ days. While Levine \& Corbet (2006) re-analyzed the RXTE/ASM monitoring data on GRO J1008-57 from 1996 -- 2005, and detected an orbital period of $\sim 248.9$ days. The independent analysis of the pulse period variations of the X-ray pulsar in GRO J1008-57 during outbursts using the BATSE data estimated an orbital period of $\sim 247.8$ days (Coe et al. 2007). Using the available RXTE, Swift, and Suzaku data of previous outbursts of GRO J1008-57, Kuehnel et al. (2013) derived an orbital period of $\sim 249.46$ days using the pulse arrival time of the X-ray pulsar.

The pulse period of GRO J1008-57 was clearly detected during its discovery in 1993 August. X-ray pulsations at a period of $\sim$ 93.587 s were detected in 20 -- 120 keV band light curves of the pulsar obtained from BATSE observations (Stollberg et al. 1993). Using ROSAT observations of the pulsar during the same X-ray outburst, $\sim 93.4$ s pulsations with a double-peaked pulse profile were detected in 0.1 -- 2.4 keV light curves (Petre \& Gehrels 1994). The double-peak pulse profile at a period of $\sim$ 93.62 s was also detected by ASCA (Shrader et al. 1999) during the 1993 outburst, and above 10 keV in the energy range of BATSE, evolved to a single-peak profile. The spin period of the X-ray pulsar GRO J1008-57 may undergo the spin-up/spin-down trend. During an outburst in 2007, a pulsation period at $\sim 93.73$ s was detected by Suzaku (Naik et al. 2011). Kuehnel et al. (2013) derive the spin period at 93.6793$\pm 0.0002$ s during 2005 outburst and at $\sim 93.7134\pm 0.0002$ s during 2007 outburst. In a recent outburst in 2012, a pulsation period at $\sim 93.648$ s was found by Swift observations (Kuehnel et al. 2013). Yamamoto et al. (2013) detected X-ray pulsations at $\sim$ 93.6257 s in GRO J1008-57 using Suzaku observation of the pulsar in 2012 November. These spin period variations are larger than the effect by its orbital period (less than $\sim 0.03$ s), suggesting that this source may undergo long-term spin period variations.

Generally the Be/X-ray pulsar systems have a magnetized neutron star with magnetic field $>10^{12}$ G according to the cyclotron resonance scattering feature (CRSF) detections in several cases as 4U 0115+63 (Li et al. 2012) and GX 304-1 (Klochkov et al. 2012). The X-ray spectra of GRO J1008-57 by different observations have not shown the significant absorption features. Shrader et al. (1999) combined the ASCA and BATSE data to obtain a broad band spectrum from 1 -- 150 keV during 1993 outburst, detecting a marginal absorption feature around 88 keV, and suggested the fundamental line energy may be about 44 keV. Suzaku observations of the pulsar during 2007 November-December X-ray outburst, however, did  not show the presence of any absorption feature in the spectrum in 0.2 -- 60 keV energy range (Naik et al. 2011).  Yamamoto et al. (2013) reported a possible absorption feature around 76 keV during the 2012 outburst by using Suzaku/PIN (15 -- 70 keV) and MAXI/GSC (60 -- 140 keV) observations. As there was no absorption feature in the spectrum at low energy (i.e. $\sim$ 40 keV), they suggested that GRO J1008-57 may be the neutron star of highest magnetic field ($\sim 6.6\times 10^{12}$ G) in known binary X-ray pulsars. Recently Kuehnel et al. (2013) found a possible absorption feature around 88 keV during the 2007 outburst with RXTE/HEXTE and Suzaku/GSO. This feature is similar to the reported feature by BATSE but inconsistent with the result by MAXI/GSC. Thus, further observations and broad X-ray spectral studies on this source are still requested.

INTEGRAL observed GRO J1008-57 during two outbursts in 2004 June and 2009 March, respectively. With a broad observational energy range from 3 -- 200 keV, INTEGRAL can well constrain the hard X-ray spectral properties of GRO J1008-57, specially providing a chance to search for the CRSFs. The INTEGRAL observations are briefly introduced in the next section \S 2. In \S 3, the spin period of GRO J1008-57 during the outburst are derived, and the long-term spin period variation is found. In addition, the orbital modulation properties of this source are also studied with available RXTE/ASM data in this section. In \S 4, the hard X-ray spectral properties in different luminosity ranges are detailedly analyzed with INTEGRAL, with main aims to search for the possible CRSFs. Finally, conclusion and discussion are presented in \S 5.

\begin{figure*}
\centering
\includegraphics[angle=0,width=14cm]{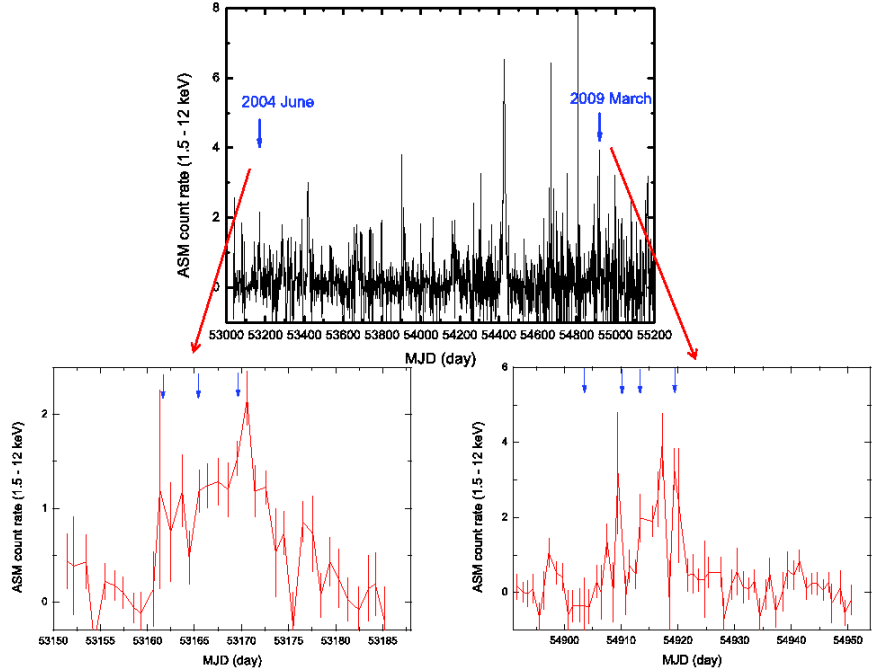}
\caption
{The INTEGRAL observation revolutions versus the RXTE/ASM count rate from 2004 -- 2009. The INTEGRAL observations are performed during two outbursts in 2004 June and 2009 March respectively (noted by arrows). }

\end{figure*}

\section{INTEGRAL Observations}

\begin{table}
%\tabletypesize{\scriptsize}
\scriptsize
\caption{\scriptsize INTEGRAL/IBIS observations of the field around GRO J1008-57 within the off-axis angle below 10$^\circ$ during two time intervals in 2004 June and 2009 March. The time intervals of observations in the revolution
number and the corresponding MJD dates, the corrected on-source
exposure time are listed. Mean IBIS count rates and the detection
significance level values in the energy range of 20 -- 70 keV were
also shown.}
% \setlength{\tabcolsep}{1.0mm}
%\tablewidth{0pt}
\begin{center}
\begin{tabular}{l c c c l}
\hline \hline
Rev. &  Obs Date (MJD)  &  IBIS rate  & Detection  &  On-source  \\
 &   &     &  significance level & time (ks) \\
\hline
201 & 53161.87 -- 53164.37 & $1.47\pm 0.06$ & 25$\sigma$ & 177 \\
202 & 53165.34 -- 53167.42 & $3.01\pm 0.07$ & 43$\sigma$ & 136 \\
203 &  53168.01 -- 53170.09 & $6.10\pm 0.07$ & 87$\sigma$ & 122 \\
783 & 54903.79 -- 54905.07 & 0.82$\pm 0.10$ & 8$\sigma$ & 100 \\
785 & 54911.05 -- 54911.99 & 0.96$\pm 0.12$ & 8$\sigma$ & 76 \\
786 & 54913.19 -- 54914.04 & $8.75\pm 0.12$ & 72$\sigma$ & 69 \\
788 & 54919.20 -- 54920.02 & $15.53\pm 0.13$ & 117$\sigma$ & 66 \\
\hline

\end{tabular}
\end{center}
\end{table}

INTEGRAL observed the hard X-ray transient GRO J1008-57 during two outbursts occurring in 2004 June and 2009 March (see Fig. 1). We mainly use the data collected with the low-energy array called IBIS-ISGRI (INTEGRAL
Soft Gamma-Ray Imager, Lebrun et al. 2003). IBIS/ISGRI has a 12' (FWHM)
angular resolution and $\sim 1$ arcmin source location accuracy in the
energy band of 15 -- 200 keV. JEM-X as the small X-ray detector aboard INTEGRAL (Lund et al. 2003) collects the lower energy
photons from 3 -- 35 keV which is used to constrain the soft X-ray band spectral properties of GRO J1008-57 combined with IBIS.

During the outburst in 2004 June, GRO J1008-57 was observed by INTEGRAL in three satellite orbital revolutions (one revolution $\sim$ 3 days), whereas during the outburst in 2009 March, INTEGRAL observations on the source were available for four revolutions. In Table 1, the information of available INTEGRAL observations used in this paper is summarized. The archival data used in our work are available from the INTEGRAL Science Data Center (ISDC).

\begin{table}
%  \tabletypesize{\scriptsize}
 \caption{The spin period of GRO 1008-57 according to different measurements}
% \setlength{\tabcolsep}{1.0mm}
%\tablewidth{0pt}
\begin{center}
\begin{tabular}{l c c l}
\hline \hline
 MJD &  Spin Period (s) & Missions & references \\
\hline
 49182  & $93.587\pm 0.015$ & BATSE & Stollberg et al. (1993) \\
 49205  & $93.621\pm 0.011$ & ASCA & Shrader et al. (1999)  \\
 50176 & $93.55\pm 0.02$ & BATSE &  Shrader et al. (1999) \\
 53166  & 93.663$\pm 0.006$ & INTEGRAL & this work \\
 53168  & 93.664$\pm 0.003$ & INTEGRAL & this work \\
 53168  & 93.668$\pm 0.001$ & INTEGRAL & Coe et al. 2007\\
 53430 &  93.6793$\pm 0.0002$ & RXTE &  Kuehnel et al. (2013) \\
 54430 &  $93.7134\pm 0.0002$ & RXTE & Kuehnel et al. (2013) \\
 54432 & 93.737$\pm 0.001$ & Suzaku & Naik et al. (2011) \\
 54912  & 93.733$\pm 0.006$ & INTEGRAL & this work \\
 54920  & 93.732$\pm 0.005$ & INTEGRAL & this work \\
 55658  & 93.727$\pm 0.001$ & RXTE & Kuehnel et al. (2013)\\
 55917  & 93.722$\pm 0.001$ & Swift  & Kuehnel et al. (2013)\\
 56245 &  93.648$\pm 0.002$ & Swift & Kuehnel et al. (2013)\\
 56251 & 93.6257$\pm 0.0005$ & Suzaku & Yamamoto et al. (2013) \\
\hline
\end{tabular}
\end{center}

\end{table}

The analysis is done with the standard INTEGRAL off-line
scientific analysis (OSA, Goldwurm et al. 2003) software, ver.
10. The bad science windows during solar activity and soon after perigee
passage are removed from our analysis. Individual pointings in each satellite revolution processed with OSA 10 are mosaicked to create sky images for the source detection. We have used the energy range of 20 -- 70 keV by IBIS for source detection and quoted source fluxes for each revolution (see Table 1). The data reduction process and background estimation are done through the standard pipeline processing (also see Bird et al. 2010).

\begin{figure*}
\centering
\includegraphics[angle=0,width=14cm]{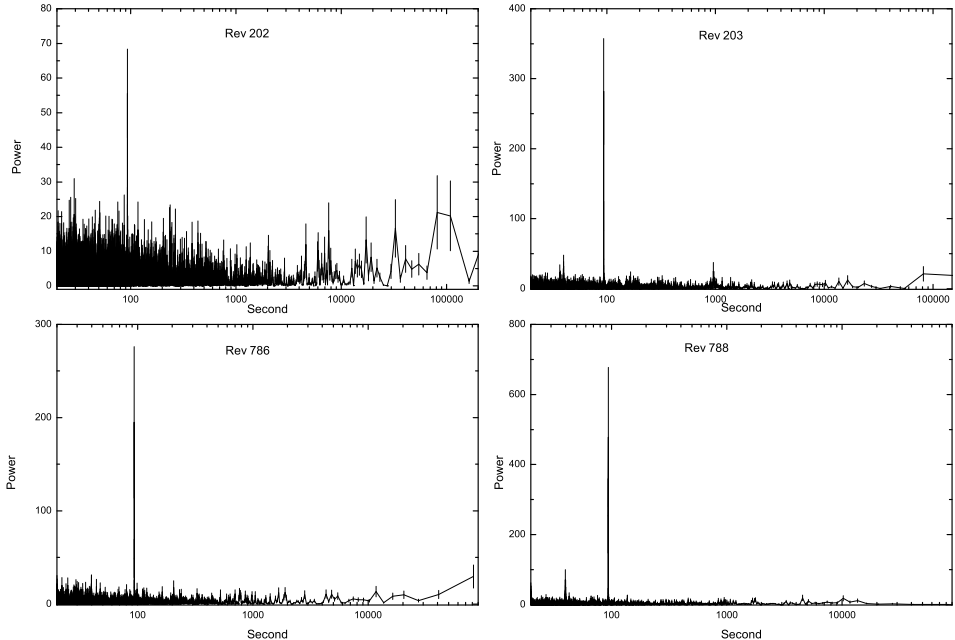}
\caption
{The power spectra for the hard X-ray light curves from 20 -- 70 keV obtained by IBIS for four revolutions: 202, 203, 786 \& 788. A significant peak at $\sim 93.7$ s is detected in all spectra.}

\end{figure*}

\begin{figure}
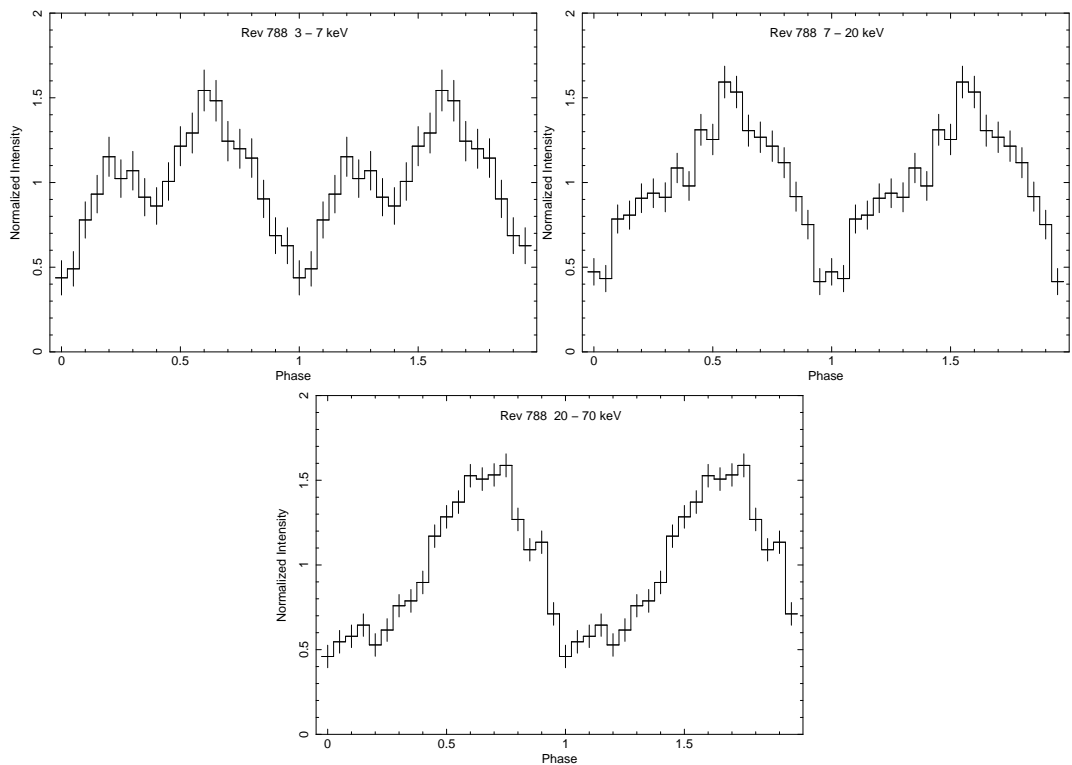

\centering
\includegraphics[angle=-90,width=7.0cm]{ms1586fig3a.eps}
\includegraphics[angle=-90,width=7.0cm]{ms1586fig3b.eps}
\includegraphics[angle=-90,width=7.0cm]{ms1586fig3c.eps}
\caption{The JEM-X and IBIS/ISGRI background subtracted light curves of GRO J1008-57 for Rev 788 folded at a pulsation period (93.732 s) in three different energy ranges: 3 -- 7 keV; 7 -- 20 keV; 20 -- 70 keV. Double peaks appear in the light curves in the lower energy range $<7$ keV; above 7 keV, a single main peak is found in hard X-ray light curves.}
\end{figure}

\begin{figure}
\centering
\includegraphics[angle=0,width=9cm]{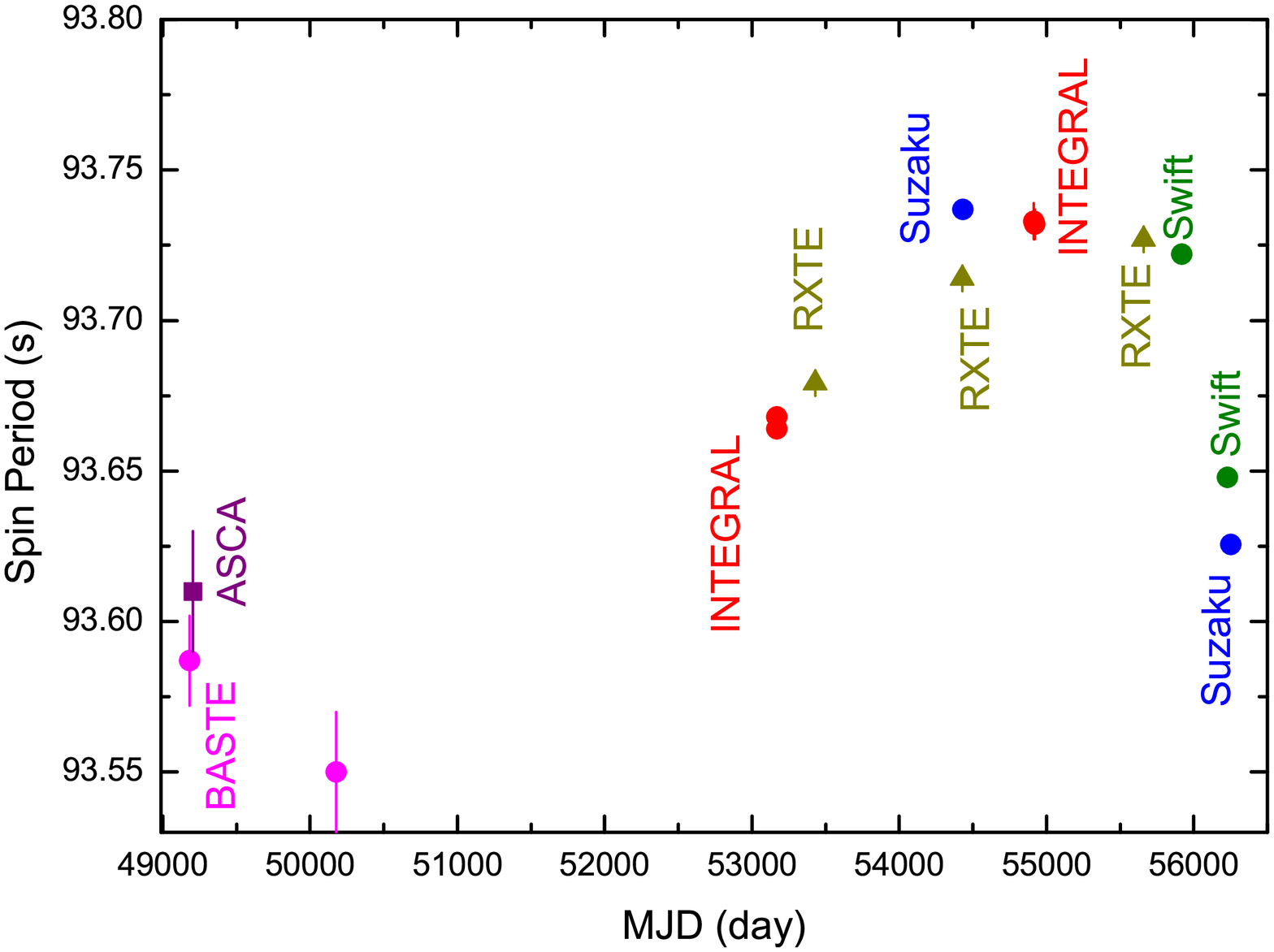}
\caption
{The spin period of the neutron star in GRO J1008-57 determined by different measurements (data points taken from Table 2). }

\end{figure}

\section{Timing Analysis}
\subsection{Long-term Spin Period Evolution}

INTEGRAL monitored GRO J1008-57 during two outbursts in 2004 and 2009. We will search for the spin period of this source in two outbursts, then possible long-term spin-period variations can be checked. The background subtracted light curves for both two detectors JEM-X and IBIS are derived for five revolutions during the outburst (IBIS detection significance level higher than 20$\sigma$): 201, 202, 203, 786 \& 788. The time resolution for each light curve is about $\sim 3$ s. Barycentric corrections have been carried out before searching for the spin period.

In Fig. 2, the power spectra of the IBIS light curves from 20 -- 70 keV are presented for four revolutions. We cannot find the strong periodic signal with the data of Rev 201. We derived the observed spin period values for the other four revolutions with the HEASOFT software package {\em efsearch} and the pulse-folding technique. To estimate  the  accurate values of spin period, the observed data were corrected for the orbital  motion of the neutron star in the binary system. Using the orbital parameters provided by Kuehnel et al. (2013), the observed spin period can be corrected to the intrinsic period for each observed time interval.  The derived spin period values of GRO J1008-57 in four revolutions are given in Table 2. The spin period of the pulsar was determined to be $\sim$ 93.66 s and 93.73 s during the 2004 June and 2009 March outbursts, respectively. With INTEGRAL observations, the long-term spin-period change trend in GRO J1008-57 is found.

The pulse profiles in 3--7 keV, 7--20 keV, and 20--70 keV ranges, obtained from JEM-X and IBIS data for Rev 788 are presented in Figure 3. The pulse profiles are found to change  with energy. At soft X-ray energy ranges ($< 7$ keV), the pulse profile appears as double-peaked whereas at high energies ($> 7$ keV), it becomes a single peaked profile.  However, the pulse fraction remains constant in all energy ranges.

In Fig. 4 and Table 2, we collected the spin period measurements obtained by different missions in the past. All spin period values are obtained during the outbursts near the neutron star periastron passage. From 1990's to 2009, GRO J1008-57 should undergo a long-term spin-down trend with spin period varying from $\sim 93.55$ s to $\sim 93.73$ s. The average spin-down rate is given as $\sim (4.1\pm 0.9)\times 10^{-5}$ s day$^{-1}$.  But after 2009 (maybe around 2009 -- 2011), the source would transfer to a long-term spin-up trend. The spin period changes from $\sim 93.73$ s in 2009 to $\sim 93.63$ s in 2012 December.

\subsection{Orbital Modulations}

\begin{figure}
\centering
\includegraphics[angle=0,width=9cm]{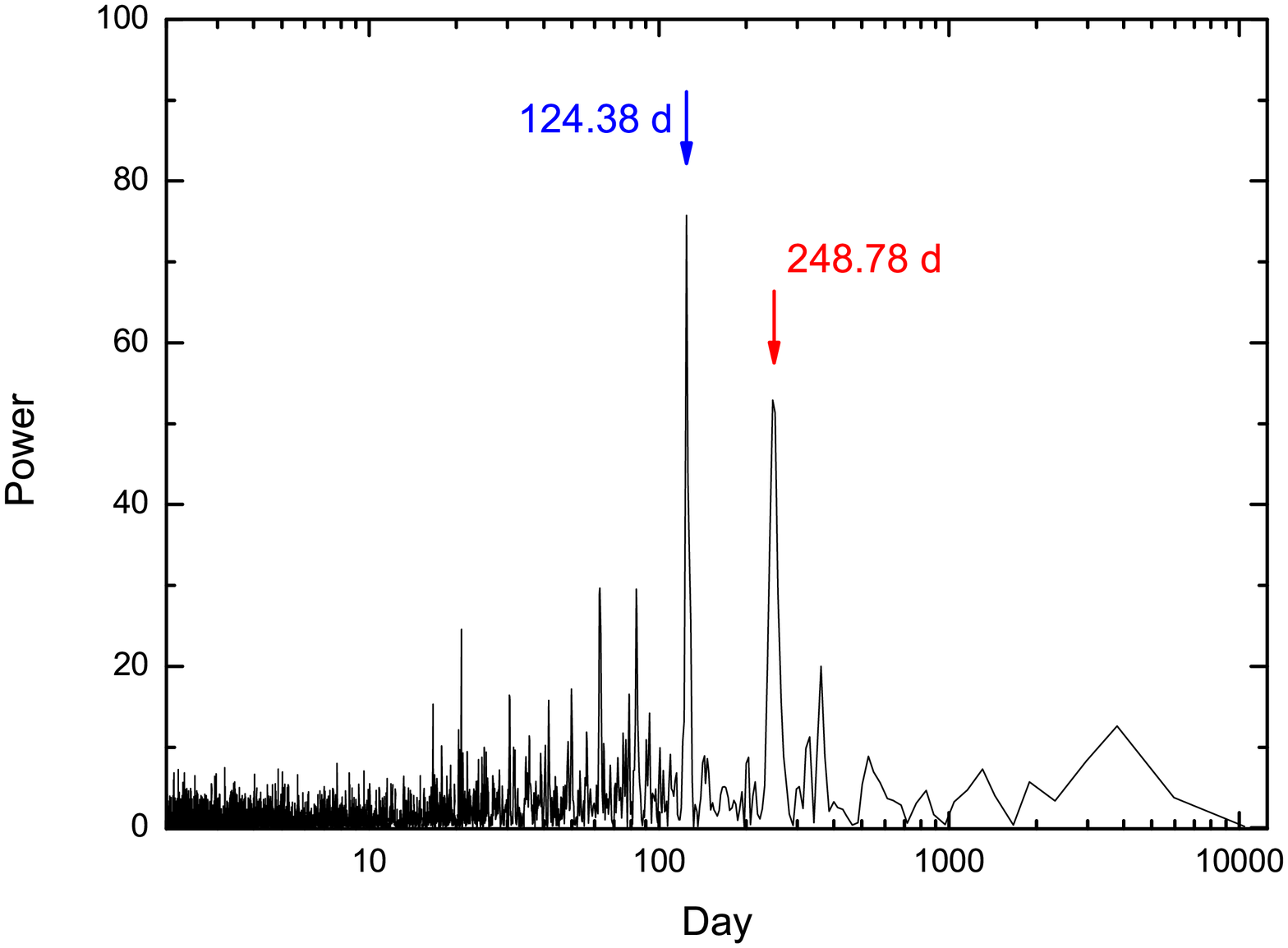}
\caption
{The power spectrum of the RXTE/ASM one-day average light curves (1.5 -- 12 keV) on the source GRO J1008-57 from 1997 -- 2011. Two significant peaks are detected, one at $\sim 248.78\pm 0.29$ day and the other at $\sim 124.38\pm 0.12$ day which is the half of the former one. The 248.78-day value is thought to be the real orbital period of GRO J1008-57 and consistent with the previous results (Coe et al. 2007; Levine \& Corbet 2006).  }
\end{figure}

\begin{figure}
\centering
\includegraphics[angle=0,width=9cm]{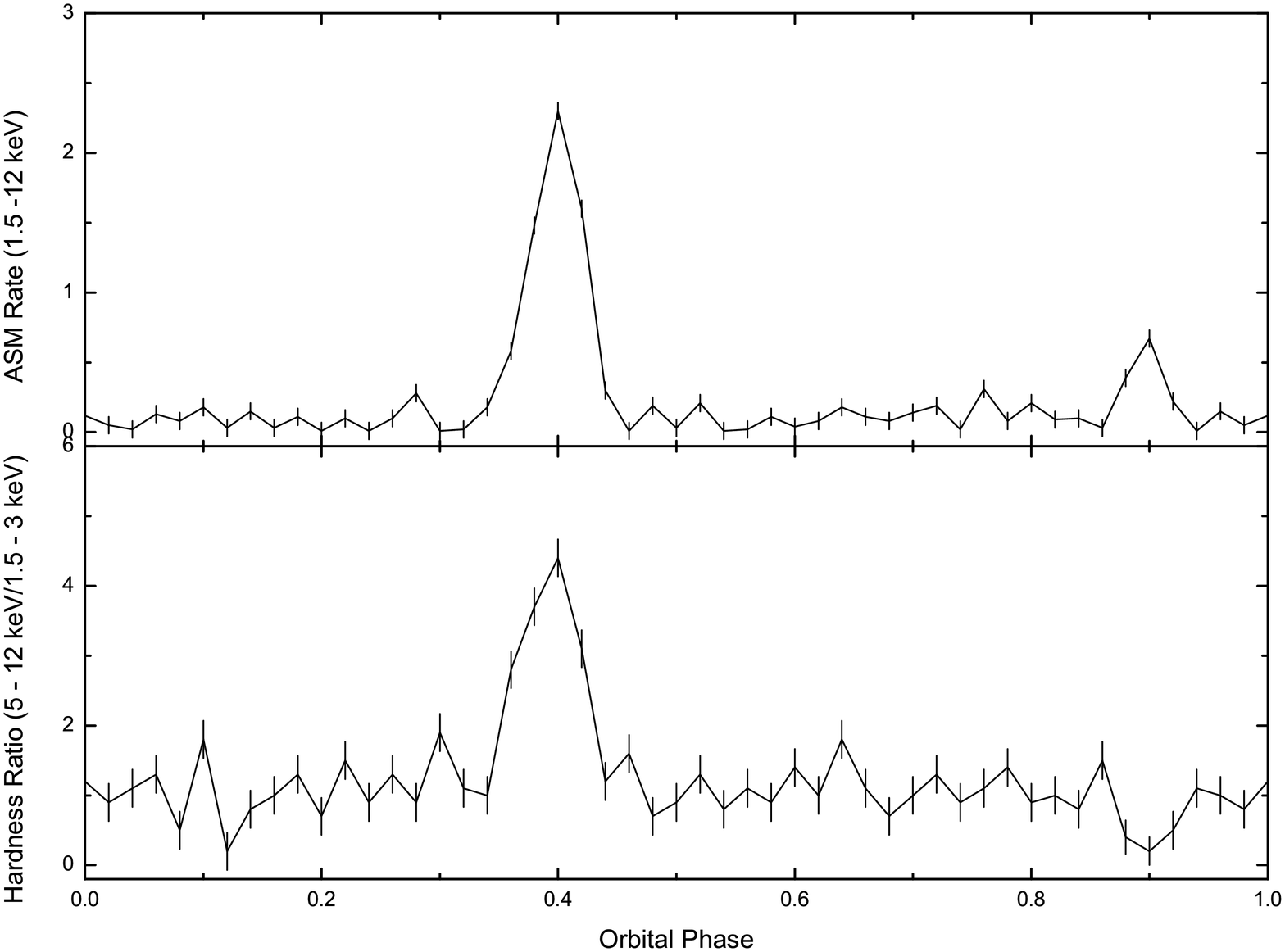}
\caption
{{\bf Top} RXTE/ASM light curve (1.5 -- 12 keV) of 4U 2206+54 folded at the orbital period of 248.78 days. The outburst peak around phase 0.4 and a second peak around phase 0.9 are found in light curve. {\bf Bottom} The hardness ratio light curve of two ASM energy bands 5 -- 12 keV/1.5 -- 3 keV folded at the orbital period of 248.78 days. The outburst peak shows the hardness ratio peak with the ratio $> 1$, but the harness ratio of the second peak is lower than 1.  }
\end{figure}

The orbital period of GRO J1008-57 has been measured in last ten years. Both the pulse arrival time technique (Coe et al. 2007; Kuehnel et al. 2012) and hard X-ray outburst recurrence (Bildsten et al. 1997; Levine \& Corbet 2006) have shown the same orbital period value around 248 days. Using RXTE/ASM monitoring data, Shrader et al. (1999) reported the orbital period of  GRO J1008-57 to be $\sim$ 136 days. Using $\sim$ 15 years of RXTE/ASM data (1996--2011),  we also study the orbital modulation characteristics of GRO J1008-57.

In Fig. 5, the power spectrum of the ASM light curve (one day averaged) from 1.5 -- 12 keV is displayed. The reported orbital period of $\sim$ 248.78 days is clearly detected. Furthermore, along with the peak at $\sim$ 248.78 days, another more significant peak appears at the period of $\sim 124.38$ days which is the half of the orbital period of GRO J1008-57. What is the physical origin of this modulation period?

We folded the ASM light curve from 1.5 -- 12 keV at the orbital period of 248.78 days which is shown in Fig. 6.  It is quite interesting that two peaks appear in the light curve from 1.5 -- 12 keV. One peak lasting about 0.1 orbital phase is the generally known type I outbursts at the neutron star periastron passage. The other peak separated by $\sim 0.5$ orbital phase from the normal outbursts is weaker than the type I outbursts in the band of 1.5 -- 12 keV, and lasts about 10 days. RXTE/ASM has three energy bands: 1.5 -- 3 keV; 3 -- 5 keV and 5 -- 12 keV. Thus we also derived the hardness ratio light curve of two bands 5 -- 12 keV/1.5 -- 3 keV, and folded it at the same orbital period. The folded hardness ratio versus orbital phases is also presented in Fig. 6. The two flux peaks have quite different spectral properties: the type I outbursts show large hardness ratio values generally higher than 1, while the second peak has hardness ratios below 1.

Thus the outbursts showing the hard spectrum become a relatively weak peak in the band of 1.5 -- 3 keV, and sometimes cannot be clearly observed. While, the second peak has a very soft spectrum which would not be observed above 5 keV. We also checked the pointing observations by INTEGRAL in the time intervals covering or near the second peaks, GRO J1008-57 cannot be detected both by JEM-X and IBIS, also suggesting a very soft spectrum which may not be significantly detected above 3 keV. So that the hard X-ray detectors like BATSE, SWIFT, INTEGRAL, can only detect the hard X-ray outburst events near the periastron, implying an orbital modulation of $\sim$ 248.8 days. While in the softer X-ray monitoring by RXTE/ASM, a soft peak component appears near the aphelion of the orbital phase, then the modulation period at 124 days could be detected.

\section{Hard X-ray Spectral Properties}

INTEGRAL observations on GRO J1008-57 reported the significant detections in seven revolutions of 2004 June and 2009 March respectively. We extract the spectrum of both JEM-X and IBIS for each revolution. Then we will fit the broad spectrum from 3 -- 200 keV combined with JEM-X and IBIS data. Generally the spectrum of accreting X-ray pulsars like GRO J1008-57 can be described by a power-law model plus a high energy exponential rolloff: $A(E)=KE^{-\Gamma}\exp(-E/E_{\rm cutoff})$. Sometimes, the spectra below 5 keV cannot be fitted well with a simple power law, then photo-electric absorption is added to fit the spectra.

The cross-calibration studies on the JEM-X and IBIS/ISGRI detectors have been done using the Crab observation data, and the calibration between JEM-X and IBIS/ISGRI can be good enough within $\sim 6\%$ (see samples in Jourdain et al. 2008 and Wang 2013). In the spectral fittings, the constant factor between JEM-X and IBIS is set to be 1. The spectral analysis software package used is XSPEC 12.6.0q.

One INTEGRAL observational revolution detected GRO J1008-57 just before the 2009 March outburst in the light curve of RXTE/ASM (Fig. 1). We define this revolution as the quiescent state when GRO J1008-57 also has a low detection significance level ($\sim 8\sigma$) by IBIS. For the other six revolutions, we define them as the outburst states. We will study the hard X-ray spectral properties in two states separately.

\subsection{Quiescent States}

\begin{table*}
%\tabletypesize{\scriptsize}
\scriptsize
\caption{Hard X-ray spectral parameters of GRO J1008-57 in seven revolutions. The flux is given in units of $10^{-9}$ erg cm$^{-2}$ s$^{-1}$ in the range of 3 -- 100 keV and the column density ($N_{\rm H}$) in units of $10^{22}$ cm$^{-2}$. For two revolutions 203 and 788, the spectral parameters are presented with two different fits: no CRSF and with CRSF. }
% \setlength{\tabcolsep}{1.0mm}
%\tablewidth{0pt}
\begin{center}
\begin{tabular}{l c c c c c c c l}
\hline \hline
Rev. Number    & $N_{\rm H}$ & $\Gamma$ & $E_{\rm cutoff}$ (keV) & $E_c$ (keV)& $Width$ (keV) & $Depth$& Flux   & $\chi^{2}\ (d.o.f)$ \\
\hline
201 & - & $1.57\pm 0.12$  &$27.4\pm 5.8$  & -&  - & - &  0.34$\pm 0.04$& 0.6739(64)  \\
202& -  & 1.33$\pm 0.07$  & $23.2\pm 2.5$  &- & - & - & $0.65\pm 0.05$ & 1.029(95)  \\
203 (no CRSF) & $4.9\pm 1.1$   & 1.29$\pm 0.07$ & $24.1\pm 1.9$ &-& - & -   &1.24$\pm 0.06$ &1.298(102)  \\
203 (CRSF) & $4.9\pm 1.1$ & 1.34$\pm 0.07$& $26.9\pm 1.9$ & 74.1$\pm 5.3$ & $<10(2\sigma)$  & $<1.9(2\sigma)$ & 1.23$\pm 0.06$ & 0.8197(99)\\
783  & -  &2.09$\pm 0.29$ & $59.7\pm 22.9$ & - & - & -  & $0.22\pm 0.06$& 0.7288(15)  \\
785  & -  & 1.07$\pm 0.33$& $13.2\pm 3.9$  &  - & - & - &$0.32\pm 0.07$& 1.068(17) \\
786&  $7.1\pm 1.5$& 1.55$\pm 0.09$ &$27.4\pm 2.8$ & -  & - & - & 1.81$\pm 0.08$ &0.7824(99) \\
788 (no CRSF)&  $5.9\pm 1.1$ & 1.38$\pm 0.06$ & $27.6\pm 1.6$ & -&  -&- & $3.11\pm 0.09$& 1.379(105) \\
788 (CRSF)& $6.3\pm 1.1$ & 1.44$\pm 0.06$ & $28.7\pm 2.1$ &73.4$\pm 3.1$ & 3.5$\pm 1.9$ & 1.7$\pm 0.8$  & $3.09\pm 0.09$& 0.8376(102) \\
\hline

\end{tabular}
\end{center}
\end{table*}

INTEGRAL observed the hard X-ray transient source GRO J1008-57 just before the X-ray outburst detected by RXTE/ASM in 2009 March (see Fig. 1) in Rev 783. Though in quiescence, GRO J1008-57 was detected by both JEM-X and IBIS. In Fig. 7, we present the spectrum of GRO J1008-57 from 3 -- 100 keV. The spectrum is fitted with a cutoff power-law model, with a cutoff energy of $\sim 59.7\pm 22.9$ keV. The large values of the cutoff energy and uncertainty suggest that the spectrum can be fitted by a simple power-law model. The power-law model gives a photon index of $\Gamma= 2.11\pm 0.24$ with a reduced $\chi^2\sim 0.8577$ (16 $d.o.f.$). The photon index of $\sim 2.1$ implies a generally harder spectrum than those of accreting X-ray pulsars. The hard X-ray flux from 3 -- 100 keV is $\sim 2.2\times 10^{-10}$ erg cm$^{-2}$ s$^{-1}$, corresponding to a luminosity of $\sim 6\times 10^{35}$ erg s$^{-1}$ assuming a distance of 5 kpc, estimated by Coe et al. (1994).

\subsection{Outburst States}

\begin{figure}
\centering
\includegraphics[angle=-90,width=7.0cm]{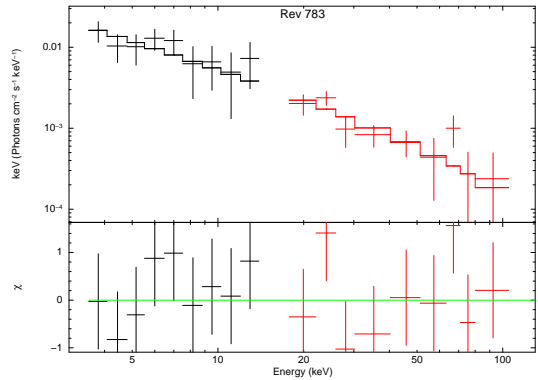}
\caption{The hard X-ray spectrum of GRO J1008-57 during the quiescent states obtained by
JEM-X and IBIS in Rev 783 (see Table 3). The spectrum is fitted by a power-law plus high energy exponential roll-off, but the simple power-law model can also describe the spectrum well.  }

\end{figure}

\begin{figure*}
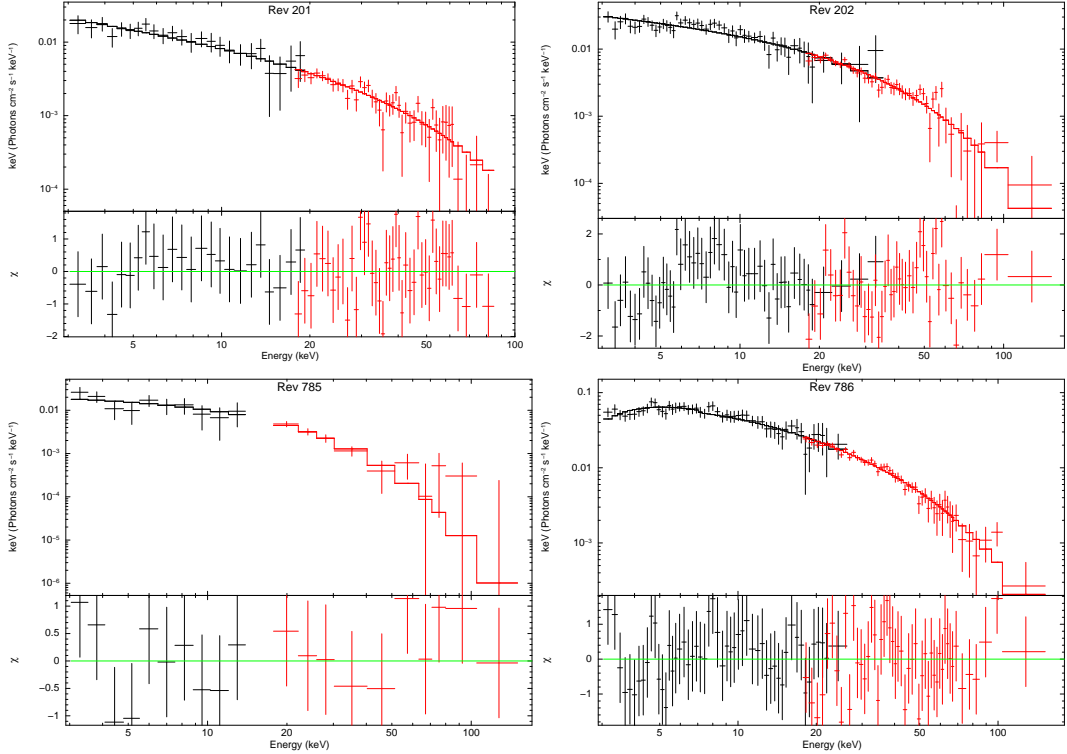

\centering
\includegraphics[angle=-90,width=7.0cm]{ms1586fig8a.eps}
\includegraphics[angle=-90,width=7.0cm]{ms1586fig8b.eps}
\includegraphics[angle=-90,width=7.0cm]{ms1586fig8c.eps}
\includegraphics[angle=-90,width=7.0cm]{ms1586fig8d.eps}
\caption{The hard X-ray spectra of GRO J1008-57 during the outburst states obtained by
JEMX and IBIS in three different observational revolutions (also see Table 3). The spectra for Revs 201, 202 \& 785 are fitted by a power-law plus high energy exponential roll-off ({\em cutoffpl}). For Rev 786, an additional photo-electric absorption is added to fit the spectra below 5 keV.}

\end{figure*}

During the outbursts, six spectra for GRO J1008-57 are extracted by INTEGRAL observations. The spectra from 3 -- 200 keV in six revolutions are well fitted by a cutoff power-law model (see Table 3, and Figs. 8 \& 9). In Rev. 785, IBIS detected the source in a low significance level ($\sim 8\sigma$), but in ASM light curve, GRO J1008-57 had gone into the outburst phase. Combined with JEM-X and IBIS, the spectral fits of GRO J1008-57 in Rev. 785 give a cutoff energy of $\sim 13$ keV which is much lower than the derived cutoff energies of $\sim 23 - 29$ keV in other five revolutions. So GRO J1008-57 shows a relatively soft spectrum in the early phase of the outburst. For three revolutions (Revs. 203, 786, \& 788) with a higher hard X-ray flux of $>\sim 10^{-9}$ erg cm$^{-2}$ s$^{-1}$ (X-ray luminosity of $\sim 3\times 10^{36}$ erg s$^{-1}$ given a distance of 5 kpc) in the range of 3 -- 100 keV (see Table 3), the additional absorption component below 5 keV is needed in the spectral fits. The large column density values of $\sim 5\times 10^{22}$ cm$^{-2}$ suggest the enhanced wind density near the neutron star periastron passage.

\begin{figure*}
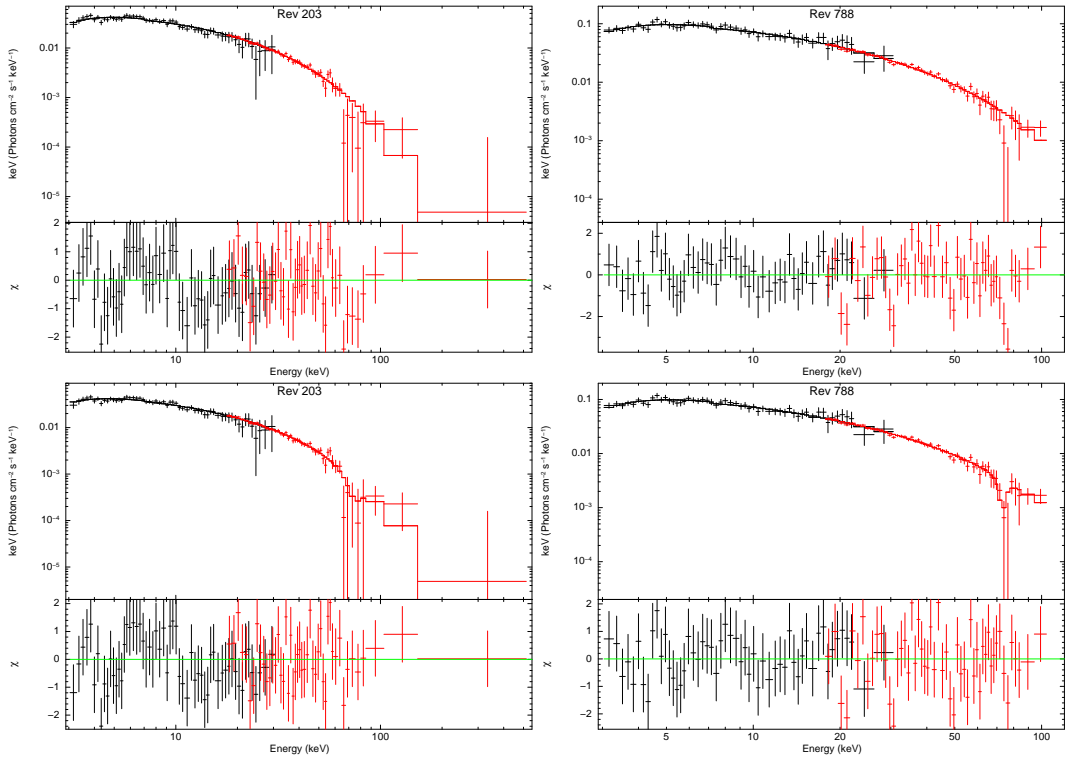

\centering
\includegraphics[angle=-90,width=7cm]{ms1586fig9a.eps}
\includegraphics[angle=-90,width=7cm]{ms1586fig9b.eps}
\includegraphics[angle=-90,width=7cm]{ms1586fig9c.eps}
\includegraphics[angle=-90,width=7cm]{ms1586fig9d.eps}
\caption{The hard X-ray spectra of GRO J1008-57 around two outburst peaks: Revs 203 and 788. The spectra are fitted an absorbed power-law plus high energy exponential roll-off (top panels) and the continuum model plus a possible cyclotron resonant scattering feature around 74 keV (bottom panels). }
\end{figure*}

We also try to search for the possible cyclotron resonance scattering features in the hard X-ray spectra of GRO J1008-57 during the outbursts. In the energy range of 10 -- 60 keV, no absorption features are discovered in the spectra of GRO J1008-57. In two observational revolutions near the light curve peaks of two outbursts (Rev 203 in 2004 June and Rev 788 in 2009 March), we find the possible absorption line features around 70 -- 80 keV in the residuals (the top two panels in Fig. 9). This absorption line feature is similar to the previously reported cyclotron absorption line feature at 76 keV by the MAXI/GSC observations (Yamamoto et al. 2013). Thus we have used the XSPEC model {\em cyclabs} to fit the possible cyclotron line features. The derived line parameters are presented in Table 3. For Rev. 203, the CRSF is not detected significantly ($<2\sigma$), the line centroid energy is derived around 74 keV, but the line width and depth can only given with the upper limits. For Rev. 788, the CRSF is determined at the energy of $\sim 73.4\pm 3.1$ keV, with a width of $\sim 2.6\pm 1.9$ keV, and a depth of $\sim 1.7\pm 0.8$. The detection significance level on this cyclotron absorption feature is not so high yet (below $3\sigma$, also comparing the reduced $\chi^2$ for the different fits by no CRSF and with CRSF in Table 3). So we only report the marginal detection of the CRSF at $\sim 74$ keV in the GRO J1008-57 during the outburst with INTEGRAL. Our results should be still consistent with the report by the MAXI/GSC observations (Yamamoto et al. 2013). Further high sensitivity and spectral resolution observations on GRO J1008-57 during the outbursts are still requested in hard X-ray bands.

\section{Summary and Discussion}

\begin{figure}
\centering
\includegraphics[angle=0,width=9cm]{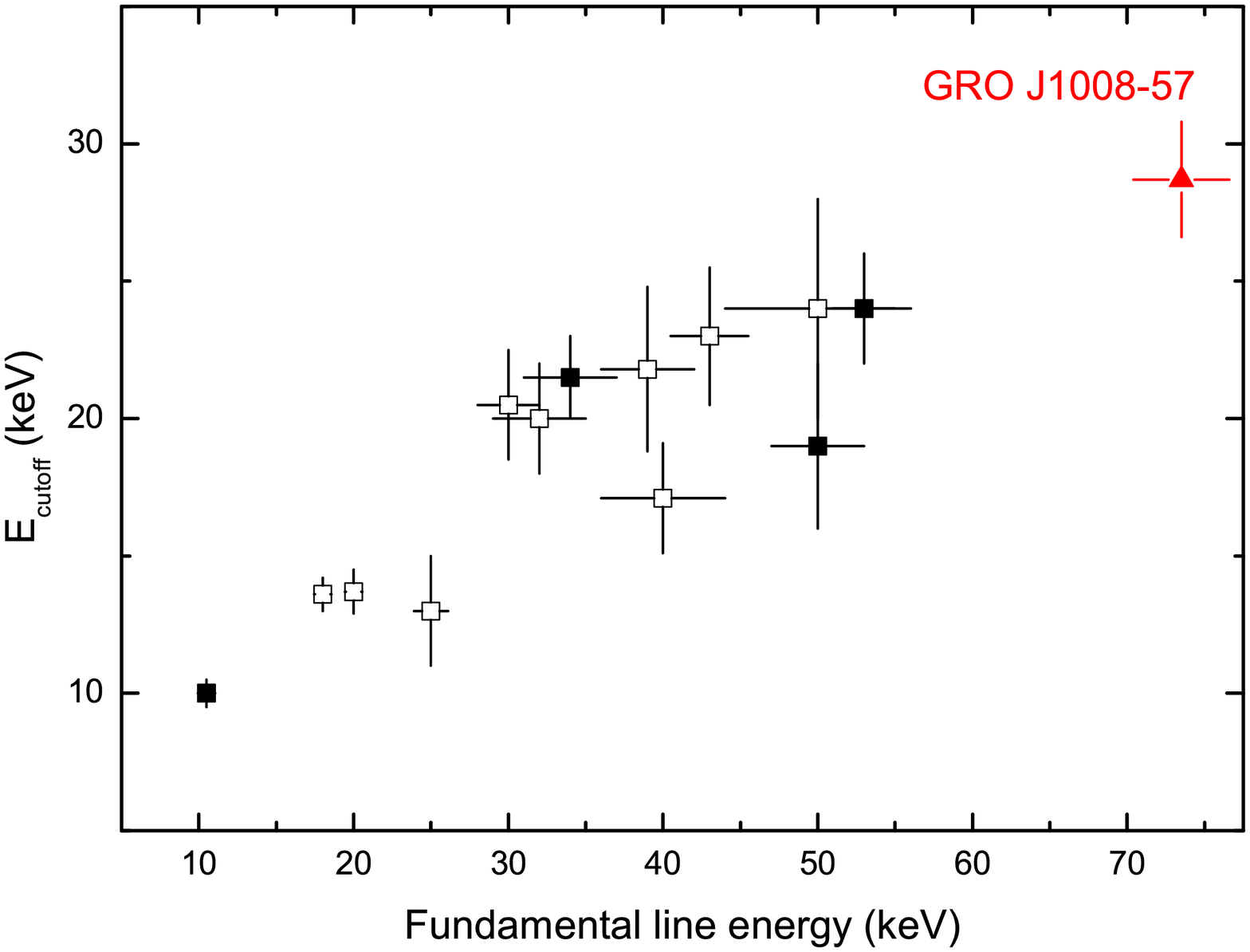}
\caption
{The centroid energy of the fundamental
cyclotron absorption line versus the cutoff energy (using a power
law with exponential high energy cutoff model) detected in accreting X-ray pulsar systems. There exists a strong correlation between the line energy and cutoff energy, suggesting that the spectral cutoff in accreting X-ray pulsars are mainly caused by the cyclotron resonance (Coburn et al. 2002). The solid squares denote the Be/X-ray pulsars, and the blank ones are other X-ray pulsar systems. These data points are collected from published literatures, i.e., Be/X-ray pulsars: 4U 0115+63 (Li et al. 2012), 1A 1118-61 (Suchy et al. 2011), GX 304-1 (Klochkov et al. 2012), XTE J1946+274 (Coburn et al. 2002); other X-ray pulsars: 4U 1909+07 (Jaisawal et al. 2013), others from Coburn et al. (2002). We also add the data point of GRO J1008-57 measured by INTEGRAL in the diagram. GRO J1008-57 follows the correlation well with the derived $E_c\sim 73.4$ keV and $E_{\rm cutoff}\sim 28.7$ keV by INTEGRAL in Rev. 788. }
\end{figure}

In this work we analyzed the spin period variations and spectral properties of the transient source GRO J1008-57 during the outbursts in 2004 June and 2009 March with INTEGRAL. We derived the spin period of GRO J1008-57 at $\sim 93.66$ s in 2004 June and $\sim 93.73$ s in 2009 March. Combined with previous spin period measurements, we find the source should undergo the long-term spin-down trend from its discovery in 1993 to 2009 with a spin-down rate of $\sim 4.1\times 10^{-5}$ s day$^{-1}$, and might transfer to a long-term spin-up trend from 2009 to now. The hard X-ray spectrum of GRO J1008-57 during the outbursts can be well described by a cutoff power-law model, with a mean photon index of $\sim 1.4$ and cutoff energies of $\sim 23 - 29$ keV. For Rev 785 when GRO J1008-57 is in the early phase of the 2009 March outburst, the spectrum is softer than those around the outburst peaks, making GRO J1008-57 a weak hard X-ray source in IBIS detections. It is also interesting that IBIS and JEM-X detected GRO J1008-57 during its quiescence just before the outburst in 2009. The quiescent spectrum can be fitted a simple power-law model with photon index of $\sim 2.1$, which is generally harder than the average spectra of accreting pulsars. In addition, above a hard X-ray flux of $\sim 10^{-9}$ erg cm$^{-2}$ s$^{-1}$ in the energy bands of 3 -- 100 keV, the additional hydrogen absorption below 5 keV is needed, implying an enhanced wind density.

The pulse profiles of GRO J1008-57 during the outbursts were studied up to 70 keV in this work, which are energy dependent: a double-peak profile in the low energy band of 3 -- 7 keV, and a single-peak profile in hard X-ray bands (above 7 keV). These energy dependent pulse profile features are also detected in other Be transient X-ray pulsars, like A0535+262 (Naik et al. 2008), 4U 0115+63 (Li et al. 2012). The double-peak pulse profile features (or dip-like structure) in soft X-ray bands can be attributed to the obscuration of matter to the radiation as suggested in other Be/X-ray pulsars. So that the observed dip-like feature in pulse profile of GRO J1008-57 will be due to the additional absorption (other than the Galactic column density) at the pulse phase. This idea is supported by our spectral analysis results of GRO J1008-57 around the burst peaks (see Table 3), the intrinsic absorption of $N_{\rm H}\sim 6\times 10^{22}$ cm$^{-2}$ is required in the spectral fits.

With RXTE/ASM data from 1996 -- 2011, we analyzed the orbital modulation properties of GRO J1008-57, and found two modulation periods at 124.38 days and 248.78 days from the light curves in bands of 1.5 -- 12 keV. The orbital period should be 248.78 days, but what are implications for the modulation period of 124.38 days? Generally, there exists an equatorial circumstellar disc around the companion star in Be/X-ray pulsar systems, the normal type I outbursts occur near the periastron passage through the disc of the Be star, and the Be/X-ray pulsar system will be in quiescence in other orbital phase. Then two possible flares could occur when the neutron star passes through the disc twice in one orbit. This idea is supported by the folded orbital light curve (Fig. 6). Two flare peaks appear in the folded light curve of the ASM data: one hard X-ray outburst peak with a harder spectrum may even disappear in the low energy band $<3$ keV; the second flare peak separated by half of orbital phase shows a very soft spectrum and cannot be observed above 5 keV. The INTEGRAL spectral analysis on GRO J1008-57 during the outburst has suggested the enhanced hydrogen absorption near the periastron passage. The hard spectrum during the outburst is mainly caused by the strong absorption in the soft X-ray bands. While the second flare near the aphelion passage with a very soft spectrum can only be detected in soft X-rays, so it is the reason that hard X-ray detectors like BATSE, SWIFT, INTEGRAL, Suzaku can only detect the 248-day orbital modulation. The suggested equatorial circumstellar disc near the neutron star aphelion passage has much lower hydrogen density than that of the periastron passage, so that the low accreting rate into the neutron star surface would produce a thermal spectrum with a typical temperature of $\sim 1$ keV and this thermal spectrum could not be strongly affected by absorption because of low hydrogen density near aphelion. This double disc-passage scenario is also suggested to explain two modulation periods found in other three high mass X-ray binaries: a Be/neutron star binary GRO J2058+42 (Corbet et al. 1997), a supergiant system 4U 1907+09 (Mashall \& Ricketts 1980) and a peculiar main-sequence companion X-ray binary 4U 2206+54 (Wang 2009). Now GRO J1008-57 becomes the second case of Be/X-ray pulsar systems with two periodic X-ray flares in one orbital phase. Why the second peak shows very soft spectrum and what is the spectral characteristics of flares near the neutron star aphelion passage? These issues need further work, specially would require the follow-up observations by soft X-ray telescopes like Chandra and XMM-Newton.

Near the outburst flux peak, the possible 74 keV absorption line feature is detected in the hard X-ray spectrum of GRO J1008-57 by INTEGRAL. The present detection is still marginal with a significance level of $\sim 3\sigma$. Our detection is also consistent with the independent measurements reported by MAXI/GSC. In addition, we find that the derived fundamental line energy and cutoff energy in GRO J1008-57 by INTEGRAL are well consistent the relationship between the fundamental line energy and cutoff energy found in accreting X-ray pulsar systems. In Fig. 10, we collected the previous observational data with cyclotron line detections and high energy cutoff measurements in accreting X-ray pulsars and plotted the diagram of fundamental line energy versus cutoff energy. The data point of GRO J1008-57 is also plotted, which is located in the upper panel of the correlation between $E_c$ and $E_{\rm cutoff}$. This consistence also implies the cyclotron line feature at $\sim 74$ keV in GRO J1008-57. So GRO J1008-57 is the magnetized neutron star of the largest fundamental cyclotron energy (highest magnetic field due to electron absorption calculation) in known accreting X-ray pulsars.

The long-term spin period variation of GRO J1008-57 is discovered from different mission observations from 1990's to 2012. From Fig. 4, a spin-down trend dominates the spin evolution of the source from 1990's to 2009. Generally, this fast long-term spin-down trend is thought to be the evidence for the propeller effect which removes the angular momentum of the neutron star by the interaction between matter and magnetosphere (Illarionov \& Sunyaev 1975; Bildsten et al. 1997). In the propeller phase of neutron star binaries, the spin-down rate can be estimated by the following formula (Illarionov \& Sunyaev 1975): \beq \dot P = {10\pi\mu^2\over GM^2R^2}, \enq where $\mu=BR^3/2$, $B$ is the surface magnetic field of neutron star, $M$ is the mass of neutron star, and $R$ is the radius of the neutron star. If we take $M=1.4$\ms, $R=10$ km, and $\dot P=4.1\times 10^{-5}$ s day$^{-1}$, the derived the surface magnetic field $B=6.2\times 10^{12}$ G. This magnitude of magnetic field is consistent with the derived strength from the cyclotron absorption line. It is quite interesting that after 2009, maybe some time around 2009 -- 2011, the neutron star in GRO J1008-57 transferred into a spin-up trend. The spin-up rate was very large, near 2$\times 10^{-4}$ s day$^{-1}$ from 2011 to 2012. What is the mechanism of the accretion torque reverse in Be/X-ray pulsars? We do not understand it yet. In some wind-fed supergiant X-ray binaries, this accretion torque reverse effect was also observed (Bildsten et al. 1997). Thus the torque reverse may be a common phenomena in neutron star binaries independent on the companion types and accretion channels. In the early of 1990's, GRO J1008-57 may undergo a spin-up trend from 1993 -- 1996 from Fig. 4, but the data points are small. Now or in future, we could still monitor the Be/X-ray pulsar GRO J1008-57 with INTEGRAL, Swift or Suzaku, studying the spin evolution, specially confirming the present spin-up trend. Studying GRO J1008-57 would help us to understand the physics of accretion torque reverse, and resolve the accreting mystery of neutron star binaries.

\begin{acknowledgements}

We are grateful to the referee for the fruitful suggestions to improve the manuscript. This paper is based on observations of
INTEGRAL, an ESA project with instrument and science data center
funded by ESA member states. We also use the archived data from RXTE/ASM in the ASM website provided by MIT. The work is supported by the
National Natural Science Foundation of China under grant 11073030.
\end{acknowledgements}

\end{document}